\documentclass[doublespacing]{elsart}

\usepackage[xdvi]{graphicx}

\usepackage{amssymb}

\begin{document}

\begin{frontmatter}



\title{Constraint on solar axions \\ from seismic solar models}


\author{Satoru Watanabe\corauthref{cor1}},
\ead{watanabe@astron.s.u-tokyo.ac.jp }
\corauth[cor1]{Corresponding author.}
\author{Hiromoto Shibahashi}

\address{Department of Astronomy, School of Science, The University of
 Tokyo, 7-3-1 Hongo, Bunkyo-ku, Tokyo 113-0033, Japan}

\begin{abstract}
 We reexamine the theoretical limit on the axion coupling to two 
 photons, $g_{a\gamma \gamma }$, in the light of a new solar model 
 (seismic solar model) and the latest solar neutrino observations, 
 the Super-Kamiokande and the Sudbury Neutrino Observatory (SNO).
 From the comparison of the theoretically expected and the measured 
 neutrino fluxes, we set a limit 
 $g_{a\gamma \gamma } < 4.0\times 10^{-10} ~\mathrm{GeV} ^{-1}$. 
 This limit based on a new procedure is about a factor of 3 improvement 
 over the previous theoretical limit and a more severe limit than the 
 solar axion experiments. Therefore this limit is the most stringent 
 limit on solar axions.  
\end{abstract}

\end{frontmatter}

\section{Introduction}
The axion is a light pseudoscalar particle introduced to solve the 
strong CP problem \cite{PQ,axion} and one of the most likely candidates 
for the dark matter. Therefore the axion is an extremely
important particle for particle physics and astronomy.

In the interior of the Sun, blackbody photons can convert into axions 
in the fluctuating Coulomb fields of the charged particles in the 
plasma,
$\gamma + (e^- ,Ze) \to (e^- ,Ze) + a$, and this reaction is called as
the Primakoff process. Large number of current and proposed
experiments are trying to detect thus-created solar 
axions \cite{avignone,moriyama,cebrian,zioutas}.

The axion interacts so weakly with other particles that it escapes 
freely from the
Sun once it is produced.
Therefore it functions as an energy-loss mechanism. 
Schlattl et
al. \cite{schlattl} calculated evolutionary solar models taking account
of this
axionic energy-loss. Comparing those models with constraints from
helioseismology and neutrino oscillations, they derived a solar limit
on the axion coupling to two 
photons $g_{a\gamma \gamma } < 10.0\times 10^{-10} ~\mathrm{GeV} ^{-1}$.

There are, however, some complaints about their approach as follows:
\begin{itemize}
\item Evolutionary solar models are constructed in the framework of
      stellar evolution, in which various assumptions about the past
      history of the Sun have to be adopted.
\item Constraints from helioseismology are not effectively utilized.
\end{itemize}
Therefore we reexamine the calculation by using seismic
solar models, which are free from evolutionary assumptions and utilize 
helioseismic constraints to the full.

\section{Seismic solar model}
\begin{figure}
\begin{center}
\includegraphics*[width=7cm]{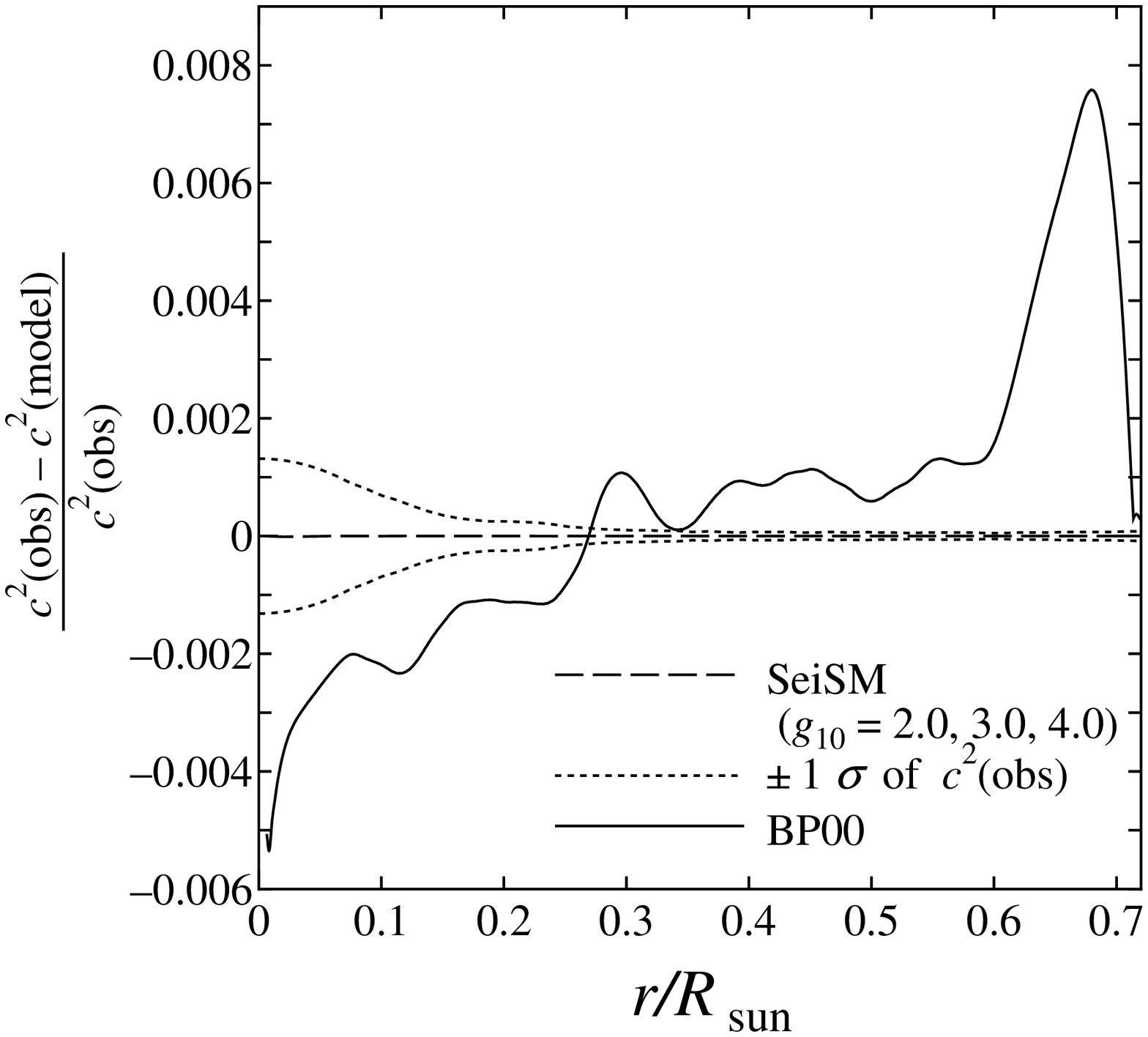}
\end{center}
\caption{Relative differences in the square of the sound-speed between
 seismic solar models with axionic energy-loss and the helioseismically 
 determined profile \protect\cite{basu}, $c^2$(obs). For a comparison, 
 the latest evolutionary model, BP00 \protect\cite{BP00}, is 
 also shown.}
\label{fig:fig1}
\end{figure}

\begin{figure}
\begin{center}
\includegraphics*[width=7cm]{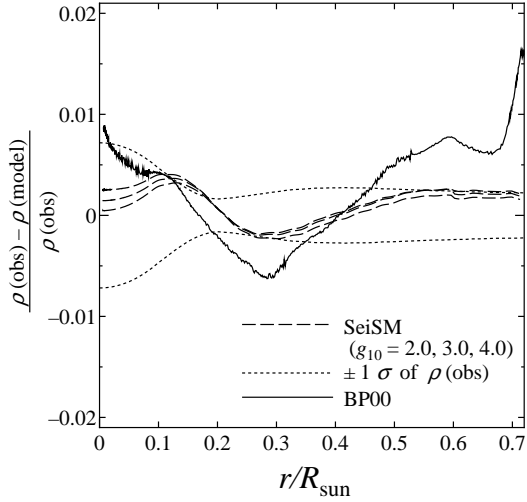}
\end{center}
\caption{Relative differences in the density. The line styles and
 references are the
 same as given in Fig. \protect\ref{fig:fig1}.}
\label{fig:fig2}
\end{figure}

Helioseismology has been successful in determining precisely the
sound-speed profile, $c(r)$, and the density profile, $\rho (r)$, in 
the Sun and the depth of the convection zone, $r_{\mathrm{conv}}$, 
based on
precise observations.
By imposing the constraints of $c_{\mathrm{obs}}(r)$, $\rho
_{\mathrm{obs}}(r)$, and $r_{\mathrm{conv}}$, we can solve the basic 
equations governing the radiative core of the Sun
(the continuity equation, the hydrostatic equation, the energy equation,
and the energy transfer equation) directly without following the
evolutionary history of the Sun. We call this model as the seismic solar
model (hereafter SeiSM) and SeiSM is described in detail in
Ref. \cite{SeiSM01}. 
SeiSM
has advantages over the conventional solar models, evolutionary models,
as follows:
\begin{itemize}
\item We can construct a model of the present-day Sun and evaluate the
      theoretically expected neutrino fluxes
      without assuming and following the evolutionary history of the 
      Sun, which cannot
      be justified directly by observations.
\item We can construct a model without
      worrying about the treatment of convection \cite{ts98}, which are 
      not
      well-described theoretically.
\item SeiSM is faithfully consistent with almost all observations
      ($c_{\mathrm{obs}}(r)$, $\rho _{\mathrm{obs}}(r)$, 
      $r_{\mathrm{conv}}$, luminosity, and the mass ratio of heavy 
      elements to hydrogen at the
      surface, which is determined spectroscopically) except
      for the neutrino fluxes, while evolutionary models are not
      necessarily so, as shown later in Fig. \ref{fig:fig1} and
      \ref{fig:fig2}.
\end{itemize}

\section{Seismic solar models with axionic energy-loss}
The axionic energy-loss rate by the Primakoff effect can be written in
the form
\begin{equation}
\varepsilon _{\mathrm{axion}} = 0.892\times 10^{-3} ~g_{10} 
^{2} ~T_7 ^7 ~\rho _2 ^{-1} ~F(\kappa ^2 )
~\mathrm{erg} ~\mathrm{g}^{-1}~\mathrm{s} ^{-1}, 
\end{equation}
\begin{eqnarray}
F(\kappa ^2 ) = & \frac{\kappa ^2 }{2\pi ^2 } & \int ^{\infty } _{0}
dx\frac{x}{\e^x - 1} \nonumber \\
                & \times & \left[\left(x^2 + \kappa ^2 
				  \right)\ln \left(1 +
\frac{x^2 }{\kappa ^2 }\right) - x^2 \right],
\end{eqnarray}
\begin{equation}
\kappa ^2 \simeq 8.28\rho _2 T_7 ^{-3} \left(3 + X\right),
\end{equation}
where $g_{10} \equiv g_{a\gamma \gamma }/10^{-10}$ $\mathrm{GeV} ^{-1}$,
$T_7 \equiv T/10^7 ~\mathrm{K}$, $\rho _2 \equiv 
\rho /10^2 ~\mathrm{g} ~\mathrm{cm} ^{-3}$, and $X$ is the mass 
fraction of
hydrogen \cite{schlattl,raffelt86,raffelt}. We calculate a series of
SeiSMs with varying $g_{a\gamma \gamma }$, which determines the amount 
of axionic
energy-loss. The model's properties are summarized in Table
\ref{tab:tab1}. 
We show the axion luminosity, $L_{\mathrm{a}}$, the temperature, $T$, 
and the 
mass fraction of heavy elements, $Z$, at the core, the mass fraction of
helium, $Y$, at the surface, and the theoretically expected
\nuc{8}{B}-neutrino flux and neutrino capture rates for the chlorine and
the gallium experiments.
In this procedure, SeiSMs are constructed so as to be
always consistent with $c_{\mathrm{obs}}(r)$ and 
$\rho _{\mathrm{obs}}(r)$ (Fig. \ref{fig:fig1} and \ref{fig:fig2}).
In addition, $Y_{\mathrm{surf}}$ of
thus-constructed SeiSMs are always in satisfactory agreement with the 
values determined directly from helioseismic inversions as 
demonstrated in Ref. \cite{SeiSM01}.


We do not try to explain how SeiSM with axionic energy-loss can be 
realized in the evolutionary process, and such an investigation is
beyond our scope. Our purpose is to clarify what can be derived if we 
construct self-consistent solar models with axionic energy-loss, which 
are faithfully 
consistent with almost all observations except for the neutrino fluxes 
and free
of severe subjective restrictions on the evolutionary history.
In our opinion, assumptions in the standard evolutionary solar models
(e.g. initial uniform chemical composition, no mass loss, and no
accretion) are subjective restriction, which cannot be justified
directly by observations.

\begin{figure}
\begin{center}
\includegraphics*[width=7cm]{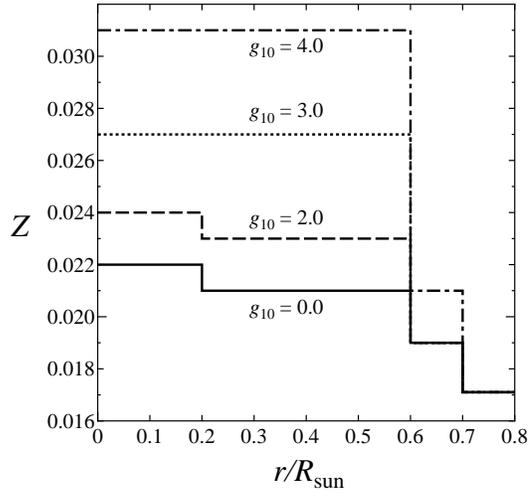}
\end{center}
\caption{$Z$-profiles of seismic solar models with different $g_{10} 
 \equiv g_{a\gamma \gamma } /10^{-10}$ GeV$^{-1}$.}
\label{fig:fig3}
\end{figure}

\begin{figure}
\begin{center}
\includegraphics*[width=7cm]{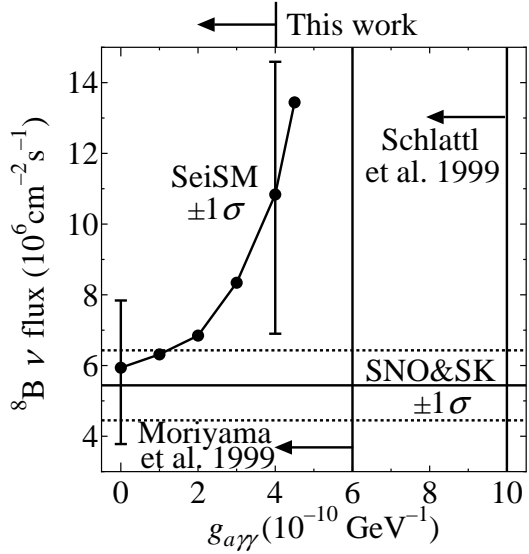}
\end{center}
\caption{The theoretically expected \nuc{8}{B}-neutrino fluxes of 
 seismic
 solar models with different $g_{a\gamma \gamma}$. For a comparison,
 the observationally determined active neutrino flux is also shown by 
 the 
 horizontal solid line with the $\pm$1 $\sigma$ error band (the dotted
 lines). Vertical lines are upper limits for $g_{a\gamma \gamma}$: this
 work, the previous theoretical limit \protect\cite{schlattl}, and the 
 most
 stringent limit from the solar axion 
 experiments \protect\cite{moriyama}.}
\label{fig:fig4}
\end{figure}

An increase in the axionic energy-loss must be compensated by an 
increase in
the nuclear energy generation, which leads to an increase in the
theoretically expected neutrino fluxes. 
An increase in nuclear energy generation is realized with increases in
$T$ and $\rho$ near the center of the Sun.
However, $\rho$ of SeiSM is
constrained by observations, so $T$ should increase particularly.
A higher $T_{\mathrm{core}}$ means a steeper temperature
gradient, which is proportional to the opacity. A higher
opacity is realized with a higher $Z$. All these trends can be seen 
in Table \ref{tab:tab1}, Fig. \ref{fig:fig3}, and later in 
Fig. \ref{fig:fig4}. 
Although $Z$-profiles shown in Fig. \ref{fig:fig3} may appear rough, it 
is
sufficient to calculate the $Z$-profile of SeiSM to the third decimal
place for discussing most of the properties of SeiSM other than the
$Z$-profile, itself \cite{SeiSM01}. 

The density constraint
and a non-uniform $Z$-profile make $T_{\mathrm{core}}$ of SeiSM very
sensitive to $g_{a \gamma \gamma}$.
Because the nuclear reaction rates are mainly controlled by
$T_{\mathrm{core}}$, the 
theoretically expected neutrino fluxes of 
SeiSM are more sensitive to $g_{a \gamma \gamma}$ than those of
evolutionary solar models with a uniform $Z$-profile \cite{schlattl}. 
We found that if we construct seismic solar models with a
uniform $Z$-profile \cite{SeiSMconst}, which are not necessarily 
consistent with
$\rho _{\mathrm{obs}} (r)$, by imposing the
constraints of $c_{\mathrm{obs}}(r)$ and $r_{\mathrm{conv}}$,
responses of the theoretically expected neutrino fluxes of such 
models to $g_{a \gamma \gamma}$ are similar to those of evolutionary
solar models with a uniform $Z$-profile \cite{schlattl}.

\section{Solar neutrinos}
Recently the Sudbury Neutrino Observatory (SNO) has announced the first
results on solar neutrinos \cite{SNO}. SNO detected 
the \nuc{8}{B}-neutrino via the charged
current (CC) reaction on deuterium and by the elastic scattering (ES) of
electrons. The CC reaction is sensitive exclusively to $\nu _{e}$,
while the ES reaction, the same reaction as the Super-Kamiokande (SK),
is sensitive to all active neutrino flavors ($\nu _{e}$, $\nu _{\mu }$,
$\nu _{\tau }$), but with weak sensitivity  to $\nu _{\mu }$ and 
$\nu _{\tau }$. If there are flavor transformations between active
neutrino flavors ($\nu _{e}$, $\nu _{\mu }$, $\nu _{\tau }$), comparison
of the neutrino flux deduced from the ES
reaction assuming no neutrino oscillations, 
$\phi ^{\mathrm{ES}}(\nu _x)$, to that measured by the CC
reaction, $\phi ^{\mathrm{CC}}(\nu _e)$, can provide the flux of 
non-electron flavor active
neutrinos, $\phi (\nu _{\mu ,~\tau})$. 
By comparing $\phi ^{\mathrm{CC}}(\nu _e)$ to the
Super-Kamiokande's precise value of 
$\phi ^{\mathrm{ES}}(\nu _x)$ \cite{S-K}, the total
flux of the active \nuc{8}{B}-neutrino, 
$\phi (\nu _{e,~\mu ,~\tau})$, is determined to be 5.44$\pm$0.99$\times
10^{6} $cm$^{-2}$ s$^{-1}$ \cite{SNO}. 
If there are also oscillations to sterile neutrinos, the total flux may 
be slightly larger. 

The neutrino flux derived in this way is consistent with the
theoretically expected neutrino flux of SeiSM with 
$g_{a\gamma \gamma } = 0$.
The relation of the observationally determined and the theoretically
expected neutrino fluxes are shown in Fig. \ref{fig:fig4}.
We estimate the uncertainty of the theoretically expected
neutrino flux by the method used in Ref. \cite{SeiSM01}.
An increase in $g_{a\gamma \gamma }$ leads to an increase in the
theoretically expected neutrino flux of SeiSM as explained in the
previous section.
Because a higher $g_{a\gamma \gamma }$ than a critical value makes the
model's neutrino flux too high to be consistent with that determined
from SNO and SK, we can limit $g_{a\gamma \gamma }$.
From the comparison of these neutrino
 fluxes, we set a limit 
$g_{a\gamma \gamma } < 4.0\times 10^{-10}$ GeV$^{-1}$.

\section{Conclusion}
We calculate a series of
seismic solar models (SeiSM) with varying the axion couplings to two
photons, $g_{a\gamma \gamma }$, which determines the amount of axionic
energy-loss. An increase in axionic energy-loss leads to increases
in the nuclear reaction rates and the expected neutrino fluxes. 
The theoretically expected \nuc{8}{B}-neutrino flux of SeiSM should be 
identical
with the total flux of the active
 neutrino flux, $\phi (\nu _{e,~\mu ,~\tau})$, which is determined
by comparing the measured neutrino flux of the Sudbury Neutrino 
Observatory
and that of the Super-Kamiokande.
From the comparison of these neutrino
 fluxes, we set a limit
$g_{a\gamma \gamma } < 4.0\times 10^{-10}$ GeV$^{-1}$. 
This limit based on a new procedure is about a factor of 3 improvement 
over
 the previous theoretical limit \cite{schlattl} and a more severe
 limit than the solar axion experiments \cite{avignone,moriyama}. 
Therefore this limit is the most stringent limit on solar axions.

\ack
This research was supported in part by a Grant-in-Aid for Scientific
Research on Priority Areas by the Japanese Ministry of
Education, Culture, Sports, Science and Technology (12047208).

\begin{table}
\caption{Seismic solar models with axionic energy-loss}
\begin{tabular}{lcccc}
\hline
$g_{a\gamma \gamma }$ $(10^{-10}$ GeV$^{-1}$)
                               &    0.0 &   3.0 &   4.0 &   4.5 \\
\hline
$L_{\mathrm{a}}/L_{\odot }$    &  0.000 & 0.019 & 0.037 & 0.050 \\
$T_{\mathrm{core}}$ (10$^7$ K) &   1.58 &  1.61 &  1.63 &  1.65 \\
$Z_{\mathrm{core}}$            &  0.022 & 0.027 & 0.031 & 0.035 \\
$Y_{\mathrm{surf}}$            &  0.240 & 0.240 & 0.240 & 0.240 \\
\hline
\multicolumn{2}{l}{Neutrino detection rates} 
                                        &       &       &       \\
\nuc{8}{B} (10$^{6}$ cm$^{-2}$ s$^{-1}$ )
                               &    5.9 &   8.3 &  10.8 &  13.4 \\
Cl (SNU$^*$)                   &    8.7 &  11.8 &  14.9 &  18.2 \\
Ga (SNU)                       &    132 &   147 &   161 &   175 \\
\hline
\end{tabular}
\vspace{6pt}\par\noindent
$^*$ A SNU is defined to be $10^{-36}$ 
interactions s$^{-1}$ per target atom.
\label{tab:tab1}
\end{table}

\end{document}